\documentclass[pra,a4paper,twocolumn,floatfix,superscriptaddress,showpacs,showkeys]{revtex4}

\usepackage[utf8]{inputenc}
\usepackage{latexsym}
\usepackage{amsmath}
\usepackage{dcolumn,bm}
\usepackage{color}
\usepackage{graphicx}
\usepackage{subfigure}
\usepackage[normalem]{ulem}
\usepackage{siunitx}
\usepackage{physics}

\newcommand{\Bc}{{\boldsymbol{\mathnormal c}}}

\newcommand{\Bl}{{\boldsymbol{\mathnormal l}}}

\newcommand{\Br}{{\boldsymbol{\mathnormal r}}}

\newcommand{\cmmnt}[1]{}

\newcommand{\superscr}[1]{\ensuremath{{}^{\rm #1}}}

\newcommand{\vp}     {\varphi}

\newcommand{\rhot}   {\ensuremath{\rho\superscr{t}}}
\newcommand{\qt}     {\ensuremath{q\superscr{t}}}

\newcommand{\rhoG}   {\ensuremath{\rho\superscr{G}}}
\newcommand{\Brho}   {\ensuremath{\boldsymbol\rho}}

\newcommand{\Q}  [2] {\ensuremath{q}^{(#1)}_{#2}}
\newcommand{\rhoZero}{\ensuremath{\rho^{(0)}}} 
\newcommand{\rhoOne} {\ensuremath{\Brho^{(1)}}} 
\newcommand{\rhoTwo} {\ensuremath{\Brho^{(2)}}}
\newcommand{\qZero}  {\Q  {0}{}}

\renewcommand{\div}{\textrm{div}}

\newcommand{\rphi}{{(\Br,\vp)}}        
        
\renewcommand{\div}{\textrm{div}}

\newcommand {\queq}[1]{(\ref{#1})}

\newcommand {\qeqs}[1]{Eqs.~\queq{#1}}
\newcommand {\qfig}[1]{Fig.~\ref{#1}}

\newcommand {\qtable}[1]{Table~\ref{#1}}

\newcommand {\beql}[1]{\begin{equation} \label{#1}}
\newcommand {\eeql}{\end{equation}}
\newcommand {\beq}{\begin{equation}}
\newcommand {\eeq}{\end{equation}}

\newcommand {\burg}{\bm{b}}

\newcommand {\etal}{\emph{et al.}}
\newcommand{\DtoC}{\emph{D2C}}

\usepackage{color}
\usepackage[normalem]{ulem} 
\newcommand{\Replace}[2]{\bgroup\noindent\textcolor{red}{\xout{#1}#2}\egroup\ignorespacesafterend}
\newcommand{\Delete} [1]{\bgroup\noindent\textcolor{red}{\xout{#1}}\egroup\ignorespacesafterend}
\newcommand{\Insert} [1]{\bgroup\noindent\textcolor{red}{#1}\egroup\ignorespacesafterend}
\newcommand{\Comment}[1]{\definecolor{Mygray}{gray}{0.50}\bgroup\color{Mygray}\noindent#1\egroup\ignorespacesafterend}

\begin{document}

\date{\today }
\title{Nanoscratching of iron: a novel approach to characterize dislocation microstructures}

\author{Nina Gunkelmann}
\email{nina.gunkelmann@fau.de}
\affiliation{Institute for Materials Simulation, 
              Department of Materials Science,
              Friedrich-Alexander University Erlangen-N\"urnberg (FAU), 
              Dr.-Mack-Str. 77, 90762 F\"urth, Germany}
\affiliation{Chair of Micromechanical Materials Modelling,
	Institute of Mechanics and Fluid Dynamics,
	Technische Universit\"at Bergakademie Freiberg (TUBAF),
	Lampadiusstr. 4,
	09596 Freiberg,
	Germany}              

\author{Iyad Alabd Alhafez}
\affiliation{Physics Department and Research Center OPTIMAS, University Kaiserslautern,
Erwin-Schr{\"o}dinger-Stra{ß}e, D-67663 Kaiserslautern, Germany}

\author{Dominik Steinberger}
\affiliation{Institute for Materials Simulation, 
              Department of Materials Science,
              Friedrich-Alexander University Erlangen-N\"urnberg (FAU), 
              Dr.-Mack-Str. 77, 90762 F\"urth, Germany}
\affiliation{Chair of Micromechanical Materials Modelling,
	Institute of Mechanics and Fluid Dynamics,
	Technische Universit\"at Bergakademie Freiberg (TUBAF),
	Lampadiusstr. 4,
	09596 Freiberg,
	Germany}

\author{Herbert M.~Urbassek}

\affiliation{Physics Department and Research Center OPTIMAS, University Kaiserslautern,
Erwin-Schr{\"o}dinger-Stra{ß}e, D-67663 Kaiserslautern, Germany}

\author{Stefan Sandfeld}
\affiliation{Institute for Materials Simulation, 
              Department of Materials Science,
              Friedrich-Alexander University Erlangen-N\"urnberg (FAU), 
              Dr.-Mack-Str. 77, 90762 F\"urth, Germany}
\affiliation{Chair of Micromechanical Materials Modelling,
	Institute of Mechanics and Fluid Dynamics,
	Technische Universit\"at Bergakademie Freiberg (TUBAF),
	Lampadiusstr. 4,
	09596 Freiberg,
	Germany}

\begin{abstract}
A new approach for characterizing the dislocation microstructure obtained from atomistic simulations is introduced, which relies on converting properties of discrete lines to continuous data. This data is represented by a number of density and density-like field variables containing detailed information about properties of the dislocation microstructure. Applying this methodology to atomistic simulations of nanoscratching in iron reveals a pronounced "length scale effect": 

With increasing scratching length the number of dislocations increases but the density of geometrically necessary dislocations remains constant resulting in decreasing shear stress. During scratching dislocations are mostly generated at the scratch front. The nucleation rate versus scratching length has an approximately
antisymmetric shape with respect to the scratch front leading to an almost constant curvature.

\end{abstract}

\pacs{
}
\maketitle

\section{Introduction}

Scratching of a surface is a standard method for evaluating  the lateral mechanical response of a material \cite{Blau2009}. A hard tip is indented into the substrate and is then moved laterally. This deformation results in pronounced plastic activity. Such scratch tests are used to determine material parameters such as the hardness and the friction coefficient. During scratching both tangential and normal hardness can be measured \cite{Wredenberg2009, Bulsara2000}. Furthermore, many scratching studies investigate the formation of defects including cracking processes and dislocation generation in single and polycrystals \cite{Caldas2011, Junge2014, Wasmer2007} and thus yield data that is equally interesting both for the plasticity community as well as for the tribology community.

Molecular dynamic (MD) simulations can be used as a tool to investigate the atomistic response of the surface during scratching. A large number of MD investigations have already been dedicated to such studies with focus on fcc materials 
\cite{Mulliah2003_NuclInstr, Mulliah2006_Nanotechnology, Komanduri2000_Wear, Zhang2012_cms}. But also scratching  of bcc substrates has been simulated, e.g.,  by Mulliah \etal{} \cite{Mulliah2003_NuclInstr} who study the depth dependence of the friction coefficient in scratching of iron, Lu \etal \cite{Lu2009_pimeJ} who use a triangular prismatic indenter and the work of Gao \etal{} focusing on dislocation evolution during scratching of Fe \cite{Gao2014,Gao2015_CMS}. 

Work hardening during scratching is a result of the evolving and interacting dislocation structure in the plastic zone around the indenter. MD simulations are able to reveal a higher degree of detail about the evolving dislocation structure than any other simulation method: they consider the trajectory of individual  atoms but require a postprocessing step to reconstruct crystal defects such as dislocation lines from the respective atom positions \cite{Stukowski2010}. Mesoscale simulation methods, such as the discrete dislocation dynamics (DDD) method (e.g., \cite{Devincre1997,Ghoniem2000,Weygand2002_MSMSE10}) keep track of the motion of each single dislocation and therefore also contain a large amount of information about the microstructure but require additional input from lower scale methods, e.g., in form of a dislocation nucleation criterion underneath the indenter. Continuum dislocation dynamics (CDD) models \citep{Hochrainer2014_JMPS,Hochrainer2015_PhilMag} describe the flow of dislocations through transport equations. By incorporating statistical averaging of discrete dislocations, CDD can capture important details about the dynamics of dislocations; at the same time it is computationally more efficient than DDD models, because the computational cost of density-based continuum methods does not scale with the number of interacting dislocations or particles. However, also CDD methods suffer from the same problem as DDD in the sense that again input from lower scale methods is required. Although MD seems to be an ideal candidate, concise methods for detailed quantitative analysis and characterization of dislocation networks as well as for scale bridging still need to be developed.
Some steps into this direction have been undertaken by Begau \etal{} \cite{Begau2012_JMPS60} who analyze dislocation density tensors of complex dislocation microstructures 
from atomistic simulations.

In order to quantitatively analyze the dislocation microstructure, we apply the recently introduced \emph{discrete-to-continuum (D2C)} method \cite{Sandfeld2015a_MSMSE23, Steinberger2016} to data obtained from atomistic simulations of nanoscratching. \emph{D2C} is a methodology for converting properties of discrete dislocation lines to continuous field data. Such data has the benefit that it is directly amenable to statistical averaging and is highly suitable for data mining. 

In section \ref{sec:MD} we first describe the molecular dynamics method of scratching followed by section \ref{sec:D2C} which introduces the relevant field variables of the CDD theory
and briefly summarizes the main features of the \emph{D2C} methodology. We analyze the field data for scratching of bcc iron for different surface orientations
and temperatures in dependence of the scratching length $L$ in section \ref{sec:results}.

\section{Molecular dynamics simulations}
\label{sec:MD}

\begin{table}
  \renewcommand{\arraystretch}{1.2}
  \centering
  \caption{The dimensions of the substrates (in the sequence of $x$, $y$ and $z$) and the number of atoms $N$ in the substrates used in our simulations.}
  \label{T0}
  \begin{tabular}{llll}
    \hline
    Plane surface  & Scratch direction  & dimensions (nm)  & $N$  \\
    \hline
    $(100)$  & $[0 \bar1 \bar1]$  & $56.1 \times 66.2 \times 24.3$  & 7796232  \\
    $(110)$  & $[001]$  & $56.5 \times 67.1 \times 26.2$  & 8603115  \\
    \hline
  \end{tabular}
\end{table}

\begin{figure}
  \centerline{\includegraphics[width=0.5\textwidth]{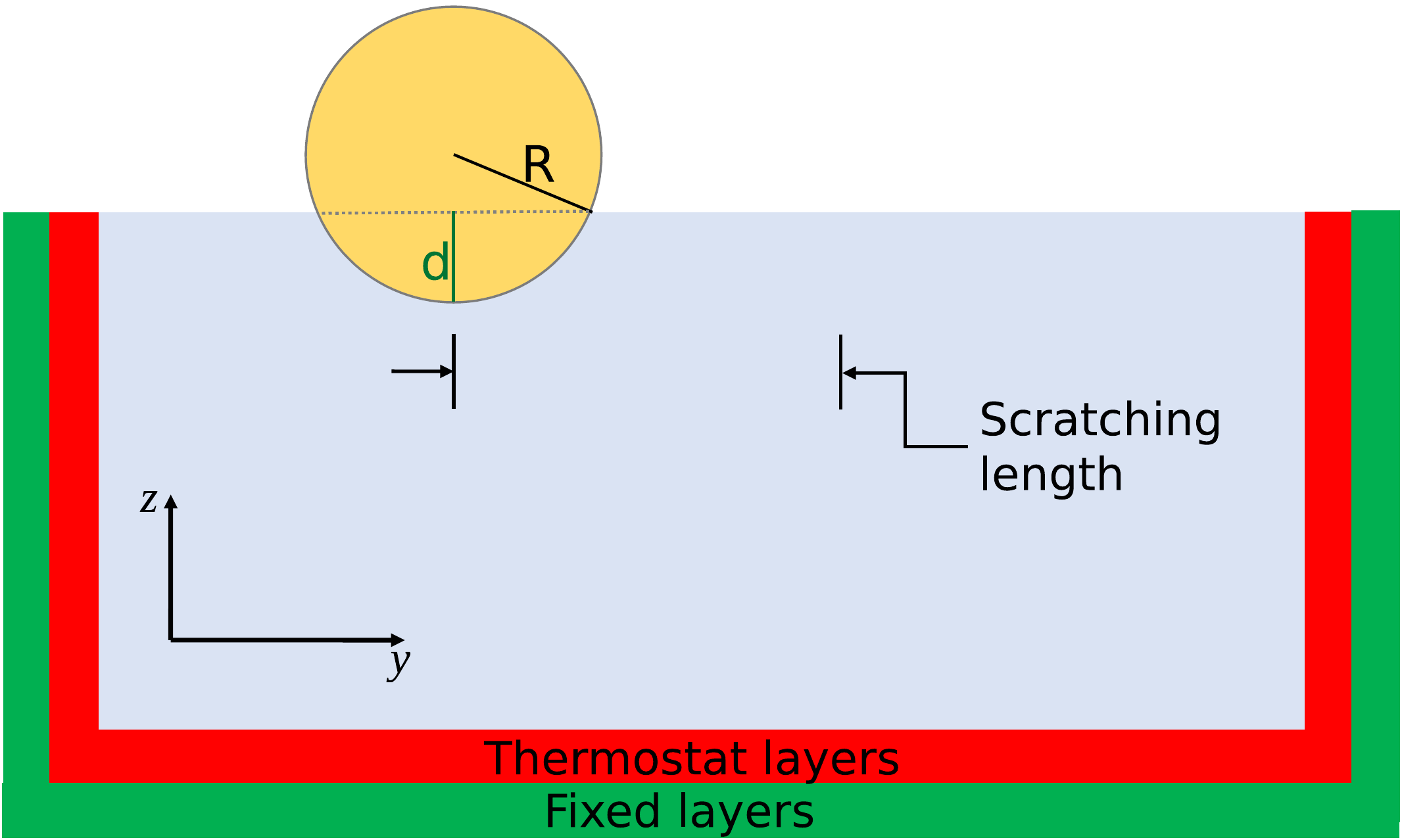}}
 \caption{Setup of the molecular dynamics simulation system, illustrating the path of the tip during scratching.} 
 \label{f_scheme}
 \end{figure}

We employ molecular dynamics simulation to study the behavior of an Fe single crystal during nano-scratching. 
The simulation system is depicted schematically in Fig.~\ref{f_scheme}, in which the configuration of the spherical tip and the Fe substrate are shown. The simulation proceeds in three steps: 
(i) The tip is indented perpendicular into the substrate surface down to a depth $d$;
(ii) the indenter moves at the indentation depth along the $y$ direction; (iii) finally, the tip is retracted from the substrate to return to its initial height above the surface. 
The scratching tip has a spherical shape with a radius of $R = \SI{10}{nm}$ and is composed of 125082 C atoms arranged in a rigid diamond lattice structure. The indenter is hollow with a thickness of approximately $0.6$ nm and moves with a constant velocity of $v = 20$ m/s. 
The depth of indentation and subsequent scratching is $d=4$~nm.
Two bcc iron single crystals were investigated. One has a (100) surface and $[0 \bar1 \bar1]$ scratching direction. The other has a (110) surface and [001] scratching direction. The crystals have lateral sizes of 56--67 nm and depths of 24--26 nm depending on the system and contain (7.8--$8.6)\times 10^6$ atoms; details are provided in \qtable{T0}. In order to prevent any transitional or rotational motion of the substrate during the simulation two atomic layers
of the substrate at the bottom as well as the lateral sides have been fixed. The next four layers are kept at a fixed temperature by a velocity-scaling thermostat. 
The substrate temperature changes from $< 1$ K to 300 K. 

The Fe-Fe interaction is described by the Mendelev potential \cite{Mendelev2003} which has a cut-off radius of 5.4 \AA. In our simulations the interaction between the diamond tip and the iron substrate is modeled by a purely repulsive potential; this is obtained from a Lennard-Jones potential describing the C-Fe interaction \cite{Banerjee2008} by prescribing a cut-off at 4.2 {\AA} at its minimum and then shifting it such that the energy and force are continuous at the cut-off radius. 

Prior to the nano-scratching simulation the Fe substrate is relaxed until all stress components reached values $< 10^{-5}$ GPa \cite{Ziegenhain2010_jap}. The indenter is placed 4.6 {\AA} above the substrate surface such that initially there is no interaction between C and Fe atoms. The scratch starts at position (0.0 0.0 0.0)~nm. 

The MD simulations were performed using the open-source LAMMPS code \cite{Plimpton1995} with a constant time step of 1 fs. {The total number of time steps simulated amounts to $0.82 \times 10^6$. Simulations require about 1.0--$1.2 \times 10^4$ cpu hours, and are performed on typically 100 nodes containing 16 cores each.}
For extracting dislocation lines from atomistic configurations we use the free software tool OVITO \cite{Stukowski2010}.

%
%

\section{The Discrete-to-Continuum (D2C) method} \label{sec:D2C}

The \DtoC{} method converts geometrical properties of discrete dislocation lines into continuous field variables \citep{Sandfeld2015a_MSMSE23}. Since geometrical dislocation lines are one-dimensional objects embedded in a three-dimensional space a point-wise comparison between two dislocation structures is difficult. In \DtoC{} dislocations are transformed into three-dimensional, continuous data by replacing discrete lines (and their geometrical properties such as line orientation or curvature) by a three-dimensional distribution function. After subdividing the three-dimensional space into averaging voxels, we can then compute volume averages of the continuous field data. 
While other authors have extracted, e.g., the Nye dislocation density tensor in a similar way \cite{Begau2012_JMPS60}, the novelty of Sandfeld's \DtoC{} approach resides in the considered field variables: \DtoC{} uses a set of continuum fields that were originally used by Hochrainer \etal{} \citep{Hochrainer2014_JMPS} and Sandfeld \etal{} \citep{Sandfeld2010_JMR26} to consistently derive a theory of continuum dislocation dynamics based on statistical averaging of systems of discrete dislocations. This CDD theory is based on a set of three variables: the total density $\rhot \equiv \rhoZero$, the first order dislocation density alignment tensor $\rhoOne$  and the curvature density $\qt \equiv \qZero$, which will be introduced in the following. 

The most general approach to obtain these variables is to add an additional orientational degree of freedom to the classical density $\rho$ (i.e., the line length per volume) and the line curvature $k$ (the inverse curvature radius). Then the density and curvature of dislocations \emph{with line orientation} $\varphi$ (i.e., the angle between the line tangent and the Burgers vector) is given by the variables $\rho\rphi$ and $k\rphi$ \footnote{These are the variables used in the original, so-called higher-dimensional CDD \citep{Hochrainer2007_PhilMag,Sandfeld2010_PhilMag90,Sandfeld2015_IJP}}, where $\Br$ denotes the spatial coordinates of a point. These two variables can easily be obtained from discrete dislocations since the line length and average line orientation in an averaging volume can be computed and also the curvature can be derived from basic geometrical relations (please refer to \citep{Hochrainer2014_JMPS} and \citep{Sandfeld2015a_MSMSE23} for further details).

Assuming that the coordinate system is aligned with the line orientation of screw and edge, the components of $\rhoOne$ are the signed screw and edge excess (geometrically necessary) dislocation densities, $\rhoOne=[\rho^{\rm s}, \rho^{\rm e}]$. They can be obtained from the field variables $\rho\rphi$ and $k\rphi$ as
\begin{eqnarray}
\label{eq:rhot}
\rhoZero(\Br)&=&\int_{0}^{2\pi} \,\!\! \rho\rphi \,\text{d}\vp \\
\label{eq:rho1}
\rhoOne(\Br)        &=&\int_{0}^{2\pi} \,\!\!\rho\rphi\Bl(\vp) \,\text{d}\vp \\
\label{eq:qt}
\qZero(\Br)  &=&\int_{0}^{2\pi} \,\!\! \rho\rphi k\rphi \,\text{d}\vp,
\end{eqnarray}
where the average line direction in a volume element is given as $\Bl(\vp):=[\cos \vp, \sin \vp]$. 
The total GND density follows as $\rhoG=|\rhoOne|$; the average line curvature can be obtained from $k=\qZero/\rhoZero$. To distinguish between edge dipoles, screw dipoles and fully isotropic statistically stored dislocation (SSD) configurations one needs to introduce an additional field variable, the second-order dislocation alignment tensor
\begin{equation}
 \rhoTwo(\Br)=\int_{0}^{2\pi} \,\!\!\rho\rphi\Bl(\vp)\otimes \Bl(\vp)\,\text{d}\vp. 
\end{equation}
The components $\rho_{11}^{(2)}$ and $\rho_{22}^{(2)}$ denote the total densities of screw and edge dislocations, respectively, and sum up to the total density,  $\rho_{11}^{(2)} + \rho_{22}^{(2)} = \rhoZero$.

The continuous fields from above can be obtained by the following strategy:
First, the domain is discretized into voxels of volume $\Delta V$. In this work, we use an edge length of the voxels of $\Delta l\sim 0.75$~nm.
The fields are then computed for each segment by extracting dislocation lines from atomistic configurations with OVITO \cite{Stukowski2010}.
Dislocation lines are approximated by cubic splines resulting in curves $\Bc$ parametrized by their arc-lengths $\varphi$. The local unit tangent vectors are $\Bl (\varphi)=\dv{\Bc}{\varphi}.$
Finally, to obtain the fields we integrate or average over all line segments within a voxel, e.g., $\rhoOne=\sum_i (\rhoOne)_i$. 
Further details, applications and a detailed mathematical description of these discrete-to-continuous (\DtoC{}) steps are presented in \citep{Sandfeld2015a_MSMSE23}.

{By comparison with the computational cost required for running MD simulations, the time required for the  \DtoC{} conversion is negligible: The conversion of a dislocation system with around 2000 segments takes less than half a second \cite{Steinberger2016_TMS}. However, using data obtained by \DtoC\ in a CDD simulation requires considerably more time. Such simulations are still much more efficient even than DDD simulations because their computational cost does not scale with the number of interacting segments. However, general CDD simulation frameworks are still under development and therefore, a detailed benchmark between MD, DDD or CDD is not possible yet.}

\section{Results} \label{sec:results}

\subsection{Continuum field description of an MD simulation}

\begin{figure*}[htbp]
	\centerline{\includegraphics{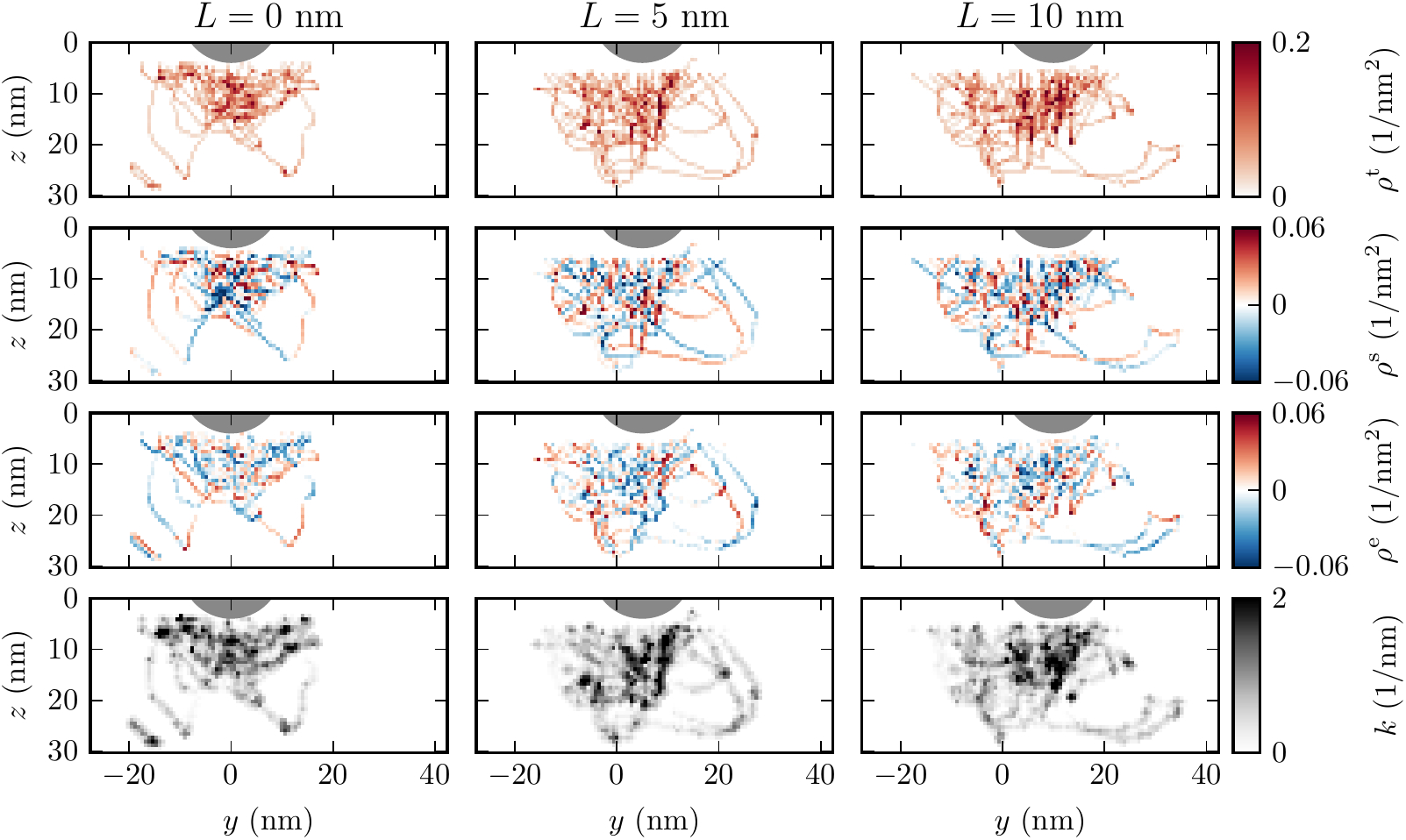}}
	\caption{CDD field variables for scratching of the (100) surface for $T<1~$K. The data is integrated perpendicular to the scratch direction (i.e., in $x$ direction). The scratching starts at $y=0.0$~nm and the indenter position is marked in gray.} 
	\label{f_scratchEvolution}
\end{figure*}

Applying the \DtoC{} method to dislocation structure extracted from a typical MD simulation for scratching of the (100) surface at a temperature of $T<1~$K results in the CDD data shown in \qfig{f_scratchEvolution}.  %
There, the left column ($L=0$ nm) shows the microstructure after the indentation prior to the lateral motion of the indenter. It can be observed that dislocations are nucleated underneath the indenter with a dislocation-free region directly under the indenter. In this region the resolved shear stress was not sufficient to trigger any nucleation events. The dislocation structure is roughly symmetrical with respect to $y=0$, typically with Burgers vectors $\burg=\frac12 \langle  111 \rangle$ and $\burg= \langle  100 \rangle$. This symmetry -- caused by the symmetrical imposed stress state -- can even be observed for the signed GND density components of screws and edges ($\rho^\text{s}$ and $\rho^\text{e}$). The GND density in \qfig{f_scratchEvolution} amounts to approximately 20~\% of the total
dislocation density $\rho^t$ (which is true more or less throughout the whole simulation). The dislocation curvature $k$ tends to be smaller for positions further away from the indenter: after a dislocation was nucleated it expands such that the radius of curvature becomes larger.

Once the scratching process starts the externally imposed stress state is no longer symmetrical. This also shows in the evolving dislocation structure where dislocations tend to glide away from their nucleation point into the scratch direction. Dislocations are mostly generated at the scratch front where both the total dislocation density and the curvature reach high values.

Upon further scratching the initially high density at the starting position of the scratch decreases while new dislocations are generated at the front. Note that in the upper half of the dislocation-rich region the magnitude of the density of screw GNDs $\rho^{\rm s}$ is slightly larger than the density of edge GNDs $\rho^{\rm e}$.

The curvature is concentrated at the scratch front where regions of high stress are expected. 
We observe that the curvature in \qfig{f_scratchEvolution} appears to be `smeared out' if compared to, e.g., the density. The reason is that the curvature is independent of the density, such that even averaging voxels with very low density will have a finite curvature value. In Steinberger \etal{} \cite{Steinberger2016} the authors have already shown that this counter-intuitive behavior does not exist if the curvature density, which is a product of density and curvature, is used.

{
 Previous work on scratching of bcc crystals using MD simulation  \cite{Mulliah2006_Nanotechnology,LGM*10,Gao2014,Gao2015_CMS,ABKU17}  provided the following results. Both $1/2\langle{111}\rangle$ and $\langle{100}\rangle$ dislocations contribute to the plastic zone created. During scratching the dislocation reorganizes, in particular by reactions of the type $1/2(111) + 1/2 (1\bar{1}\bar{1}) \to (100)$. Such reactions reduce the dislocation density in the middle of the groove. Most dislocation activity occurs in the scratch front. Our present study corroborates these results.
}

\begin{figure}[htbp]
	\centerline{\includegraphics[width=0.5\textwidth]{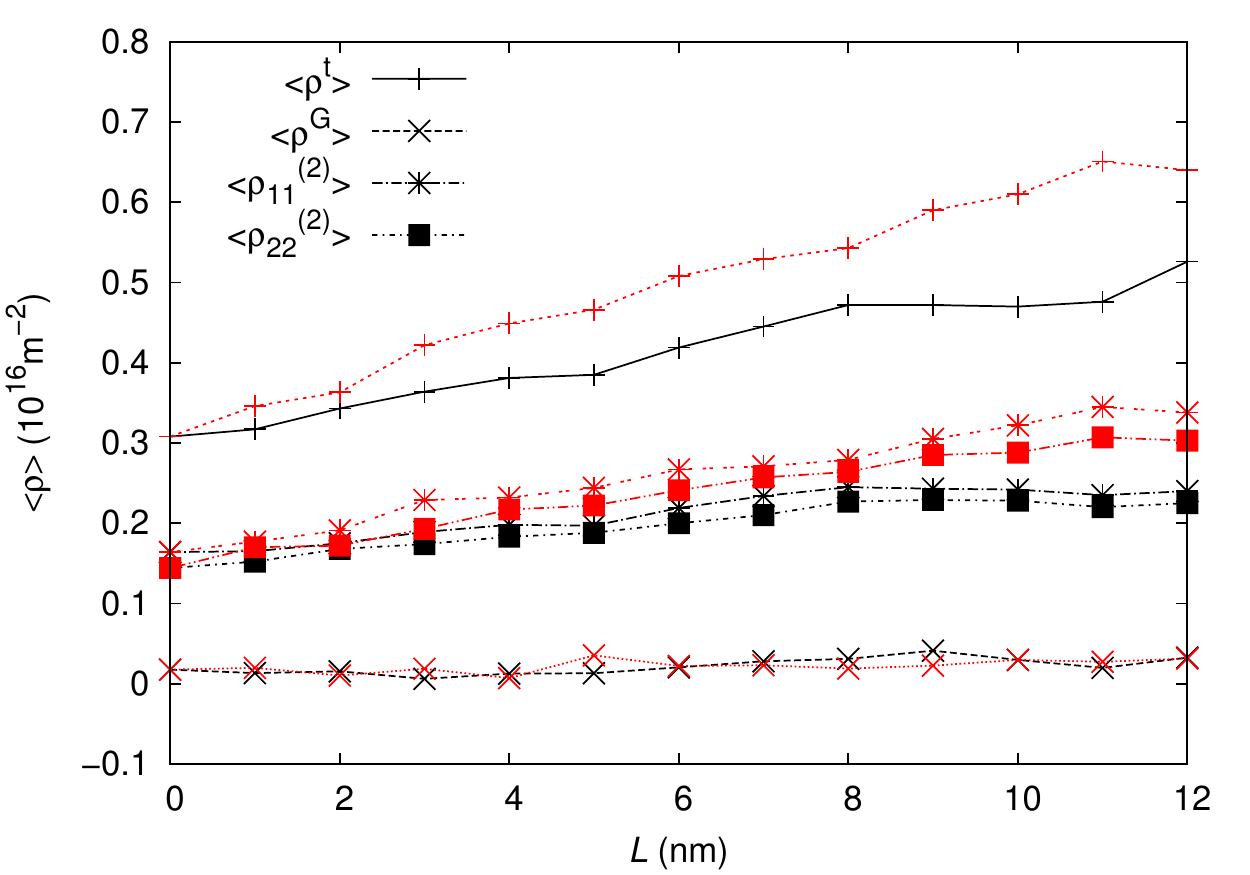}}
	\caption{Globally averaged values of various density fields versus scratching length $L$ for $T<1~$K. Data for scratching the (100) and (110) surface are shown in black and red, respectively.} 
	\label{f_dens}
\end{figure}

{
\begin{figure}[htbp]
	\centerline{\includegraphics[width=0.5\textwidth]{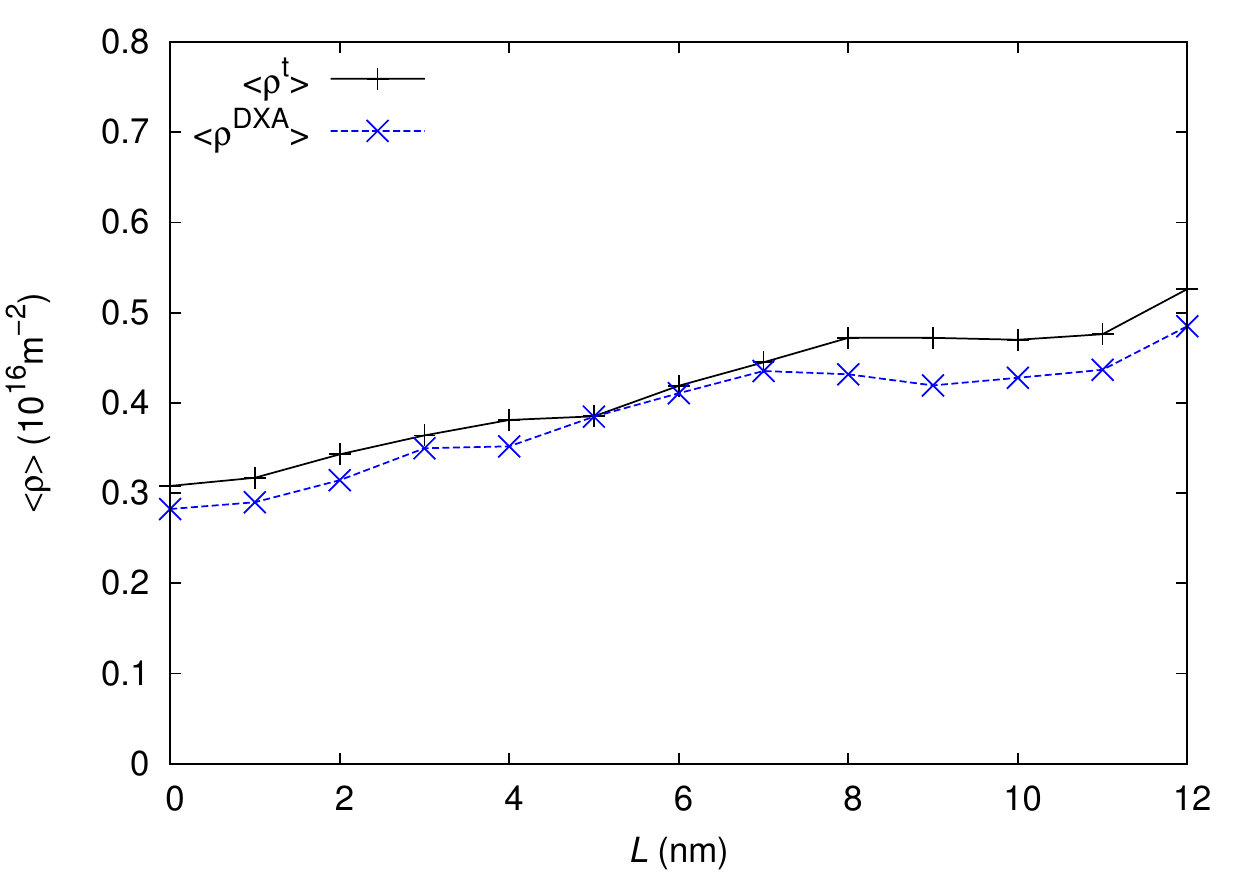}}
	\caption{{Globally averaged values of the total dislocation density versus scratching length $L$ for $T<1~$K for scratching of the (100) surface. The data obtained by DXA using a smoothing level of 1 and a point separation value of 2.5 $\langle \rho^{\text{DXA}} \rangle$ is compared to the field data.}} 
	\label{f_densComp}
\end{figure}}

\qfig{f_dens} displays the globally averaged values of the total dislocation density $\langle \rhot \rangle$, the GND density $\langle \rhoG \rangle$ and the components of the second-order dislocation alignment tensor, $\langle \rho_{11}^{(2)} \rangle$ and $\langle \rho_{22}^{(2)}\rangle$, versus scratch length. The dislocation density $\langle \rhot \rangle$ increases with scratching length for both orientations, while the number of GND dislocations $\langle \rhoG \rangle$ stays at very small values for all scratch lengths $L$. Comparing this value to the total screw and total edge density,  $\langle \rho_{11}^{(2)} \rangle$ and $\langle \rho_{22}^{(2)} \rangle$, we find that on the scale of the system all dislocations are statistically stored dislocations (SSDs). However, screw dislocations can be found more frequently than edge dislocations which shows in the fact that the component $\langle \rho_{11}^{(2)} \rangle$ is slightly higher than $\langle \rho_{22}^{(2)} \rangle$. This is in agreement with the fact that in bcc metals mainly the screw dislocations contribute to strengthening because of their small mobility. 

{To evaluate the performance of our method we compare in \qfig{f_densComp} the total dislocation density $\langle \rhot \rangle$ to the dislocation density $\langle \rho^{\text{DXA}} \rangle$ obtained by DXA. 	
The agreement between the DXA result $\langle \rho^{\text{DXA}} \rangle$ and the density obtained by {D2C} is generally very good with a relative error which in most regions is significantly below $10\%$;
errors must be attributed to D2C's generic spline approximation of the DXA polygons. These errors are dependent on the number and distance of the polygon support points, which is the reason for the variation in the relative error during microstructure evolution. Further developments of D2C will reduce the error by a spline approximation that is directly tailored to the underlying DXA algorithm. 

As a results we find that the dislocation density increases roughly linearly
with scratching length. A similar result was also found by Gao \etal~\cite{Gao2014} who studied nanoindentation and nanoscratching of iron. They found that the total dislocation 
length increases steadily until it reaches a maximum where re-organization of the dislocation network starts.}

\begin{figure}[htbp]
  \centerline{\includegraphics[width=0.5\textwidth]{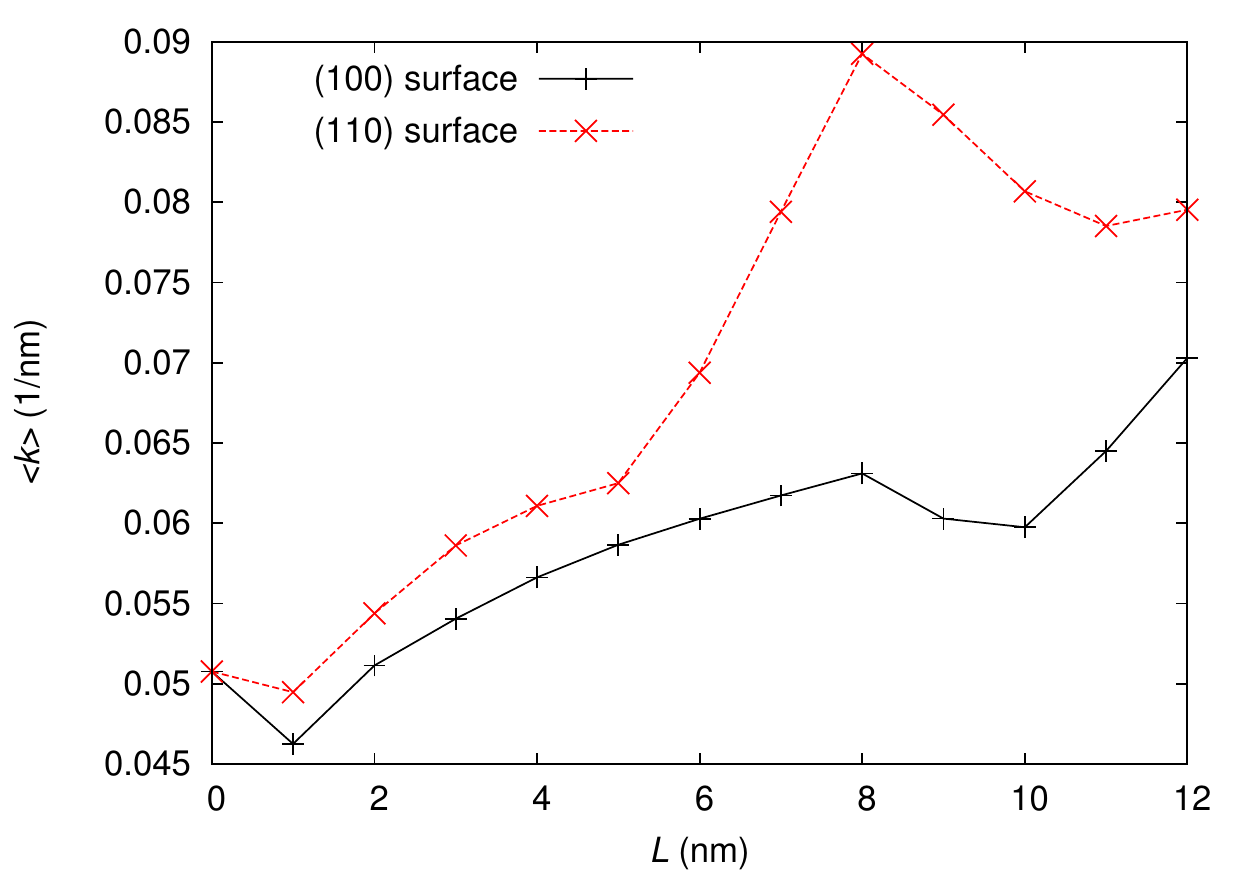}}
 \caption{Globally averaged curvature versus scratching length $L$ for $T<1~$K. Data for scratching the (100) and (110) surface are shown in black and red, respectively.} 
 \label{f_curv}
 \end{figure}

The average curvature only slightly increases with scratching length (\qfig{f_curv}) for scratch of the (110) surface, where the average radius of curvature decreases from approximately $1/(\SI{0.1}{nm^{-1}})=\SI{10}{nm}$ to around $\ 1/(\SI{0.16}{nm^{-1}})\approx\SI{6}{nm}$. It decreases less for (100) scratching. 

In accordance with \cite{Gao2015_CMS} the difference in dislocation density for different crystallographic orientations can be attributed to the crystalline anisotropy of the scratch process. We assume that this is also the reason for the difference in the evolution of the average curvature for the two different surface orientations.

%
%
%
%

 \subsection{Nucleation rate}

Upon volume integration the CDD field variables $\rhot$ and $\qt$ give access to the total line length $R$ and the total number of dislocation loops $Q$. The rate of change of these variables allows to analyze the nucleation behavior, where in particular $Q$ gives information about the generation of new dislocations loops, while a stronger increase of $R$ is rather a consequence of such increased number of dislocations. Therefore, we monitor the rate of line length production $\dot R(y)$ and the loop nucleation rate $\dot Q(y)$, 
\begin{eqnarray}
\dot R(y)=\int_{x,z} \dot{\rho}^{\text{t}} (x,y,z) \,\ \text{d}x\text{d}z \label{eq:R} \\
\dot Q(y)=2\pi^ {-1}\int_{x,z}\dot q(x,y,z) \,\ \text{d}x\text{d}z,
\label{eq:Q}
\end{eqnarray}
where the time derivative of $\rhot$ and $\qt$ in \qeqs{eq:R} and (\ref{eq:Q}) can be approximated using an forward finite difference scheme in time such that
\begin{eqnarray}
\dot{\rho}^{\text{t}}_{n+1}&=&(\rhot_{n+1}-\rhot_n)/\Delta t\\
\dot q_{n+1}&=&(q_{n+1}-q_n)/\Delta t.
\end{eqnarray}  
Here $\Delta t=t_{n+1}-t_n$ is the time step between two discrete times $t_{n}$ and $t_{n+1}$.
The $\rhot_i$ and $\qt_i$ denote the values of the respective fields at times $t_i$. {These fields were extracted from the atomistic data using \DtoC{}.}

\begin{figure*}[htbp]
	\centerline{\includegraphics{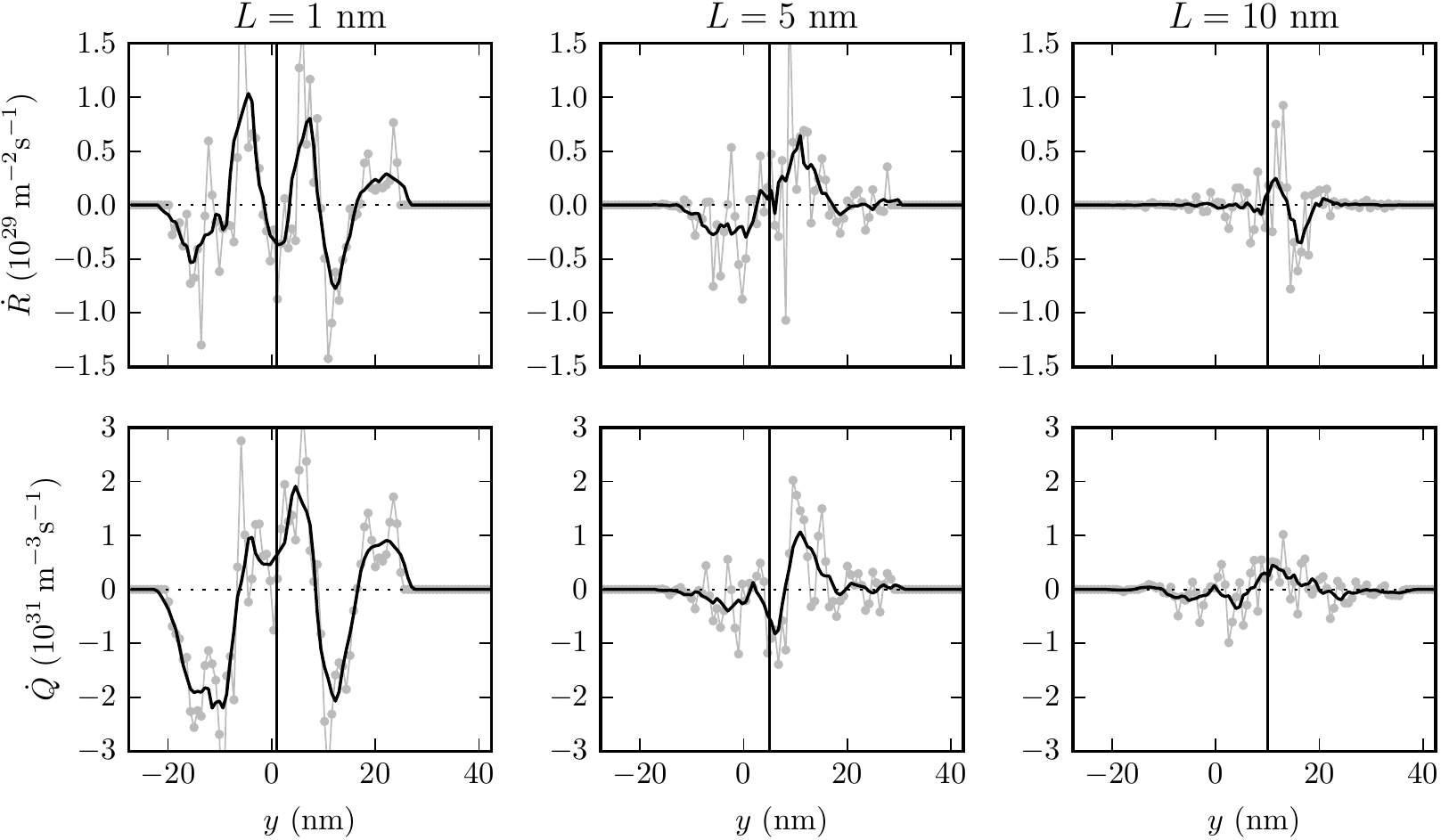}}
	\caption{Nucleation rate $\dot R(y)$ [$\dot Q(y)$] versus $y$ for $T<1~$K at $t=50$, $t=250$~fs and $t=500$~fs for scratch of the (100) surface.
		Curves are smoothed by moving averages with an averaging window size of seven data points corresponding to a length of 6~nm. The vertical lines denote
		the position of the indenter.} 
	\label{f_rhoDot}
\end{figure*}
From \qfig{f_rhoDot} we can see that the dislocation nucleation rate $\dot R(y)$ has highest values right of the scratch front (i.e., at $y\sim14$~nm for $L=10$~nm).
Except for the initial state, we always observe an approximately anti-symmetric shape of the dislocation curvature nucleation rate $\dot Q(y)$ which is positive towards the scratch front and negative at/behind the indenter which shows that dislocations are nucleated at the scratch front and removed behind the indenter (e.g., by annihilating or by flowing away). 
For $L=5$~nm the positive branch of the curves dominates and the generation of dislocations outweighs the annihilation processes.
At later times the maximum nucleation rates decrease strongly, and the generation of dislocation line length balances the removal of line length. This is consistent with the snapshots in \qfig{f_scratchEvolution} where we also observe that the dislocation structure tends to a stationary state in regions behind the nucleation front. The interplay between the nucleation rate of density and curvature density can be explained as follows: A higher loop nucleation rate $\dot Q$ results in more dislocation segments/loops, which in turn result in an accelerated density production. This explains the large positive peaks for $\dot R$ at $L=5$~nm and $L=10$~nm. The negative dip in $\dot R$ (which is most pronounced for a scratching length of $L=10$~nm with maximum at $\approx 18$~nm) indicates a loss of dislocations due to mutual annihilation.

The antisymmetric shape of the nucleation rate $\dot R(q)$ at larger scratching lengths for the (100) orientation explains why the average density $\langle \rhot \rangle$ 
remains constant with increasing scratch length (see \qfig{f_dens}): The annihilation of dislocations lines balances their generation and thus the total line length becomes independent of time.

Note that the curvature is a field variable which is directly connected to the line tension due to dislocation self-interactions. A simple line tension approximation reads \cite{Sandfeld2015_IJP}:
\begin{equation}
 \tau^\text{lt}={T_s \mu b k}. 
 \label{eq:LineTension}
\end{equation}
where $\mu$ is the shear modulus and the (orientation dependent) strength of the interaction is governed by the constant $T_s \in [0.5,1]$. 
From our scratch simulation we find that the influence of the line tension becomes -- on average -- larger with increasing scratch length (\qfig{f_curv}), while spatially the main contribution is located directly underneath the indenter (fourth row in \qfig{f_scratchEvolution}) where also the highest loop nucleation rate exists.\\

\subsection{Von Mises stress}
 
For analyzing the driving stress for plastic activity we use the von Mises stress, which is a suitable quantity to describe the deformation and failure response of scratch and indentation tests, as shown in \cite{Pelletier2008,Zhao2014, Pane2006}.
Two stress states with equal distortion energy have the same von Mises stress. 
According to the von Mises criterion yielding occurs once the deformation energy equals the deformation energy at yield in simple compression or pure shear \cite{Bhushan2013}. 
A material is found to start yielding when its von Mises stress reaches a critical value called yield strength.  Using the components $p_{ij}$ of the stress tensor for each atom, the von Mises stress per atom is defined as
\begin{widetext}
\begin{equation*}
\sigma _{\rm vM}=\sqrt{\frac{1}{2}\left[ \left( p _{xx}-p_{yy}\right)^{2} \\+\left( p _{xx}-p _{zz}\right) ^{2}+\left( p _{zz}-p_{yy}\right) ^{2}+6\cdot \left( p _{xy}^{2}+p _{xz}^{2}+p_{yz}^{2}\right) \right] }.
\end{equation*}
\end{widetext}
To evaluate the von Mises stress we average $\sigma _{\rm vM}$ under the indenter considering atoms in a cylinder of radius $r=8$~nm  with its axis along the $z$ axis for $z<9$~nm for different scratching lengths and
plot the averaged values versus dislocation density $\langle\rhot\rangle$. 
Note that we average only over disordered atoms belonging to the surface, to point and line defects. The atom structure was detected by OVITO \cite{Stukowski2010}. 
\begin{figure}
	\subfigure[] {\centerline{\includegraphics[width=0.5\textwidth]{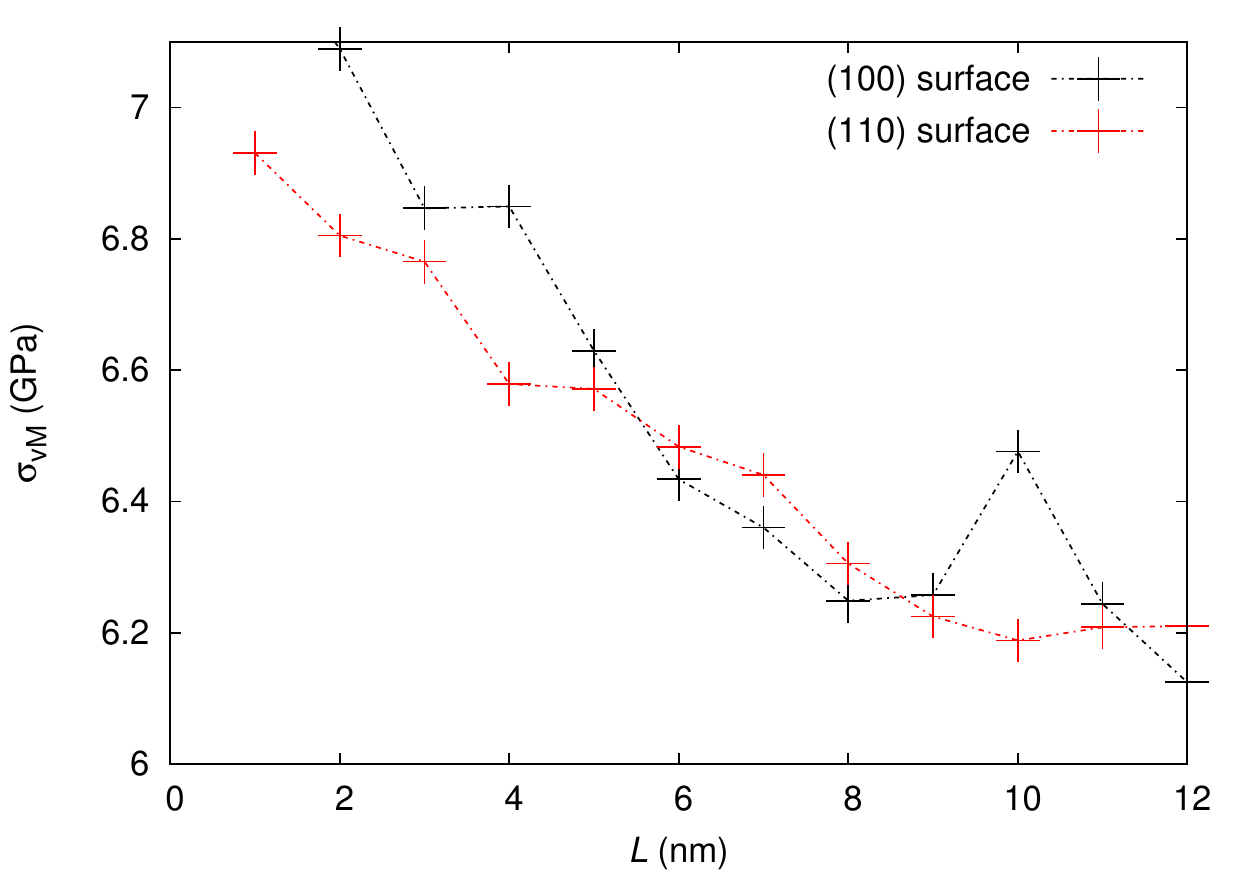}}}
	\subfigure[] {\centerline{\includegraphics[width=0.5\textwidth]{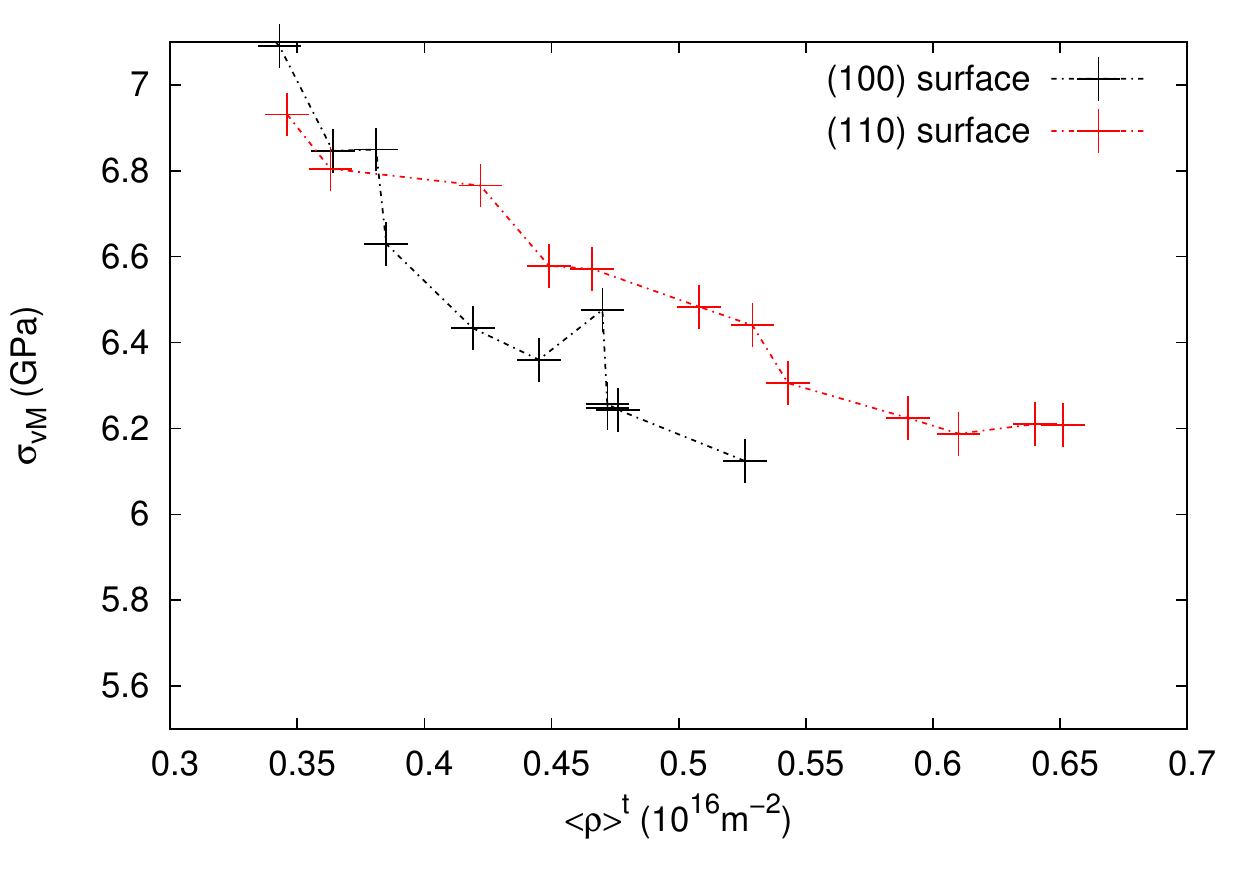}}}
	\caption{{Von Mises stress $\sigma_{\rm{vM}}$ representing the driving stress for plastic activity versus (a) scratching length $L$ and (b) total dislocation density $\langle \rhot \rangle$ for $T<1~$K.  Scratch of the (100) and (110) surface is marked in black and red, respectively.}}
	\label{f_stress}
\end{figure}
From \qfig{f_stress} we see that the von Mises stress under the indenter decreases with increasing scratching length $L$ and  dislocation density $\rhot$. We observe a reversed size effect typical for scratching with a spherical indenter: With increasing scratching length the number of
statistically stored dislocations increases but the number of GNDs $\langle\rhoG\rangle$ stays constant \cite{Silberschmidt2015}. The shear stress, therefore, displays a reverse dependence
on scratching length and dislocation density. 

Note that the normal and tangential hardness also decrease with increasing scratching length \cite{Gao2014}, which is a direct consequence of the definition of hardness, which is experimentally measured as the ratio of applied forces to contact areas. Thus the normal and tangential hardness values measure the response of the material to normal and tangential forces. The two hardness values may differ from each other due to crystalline anisotropy effects of plasticity and different loading conditions \cite{Gao2015_CMS}. Furthermore, Gao \etal{} showed in \cite{Gao2014} that the material appears to soften with increasing scratching length. Due to the observed reverse dependence of the shear stress on scratching length and density our simulations exhibit a similar behavior.

\subsection{Averaging over repeated MD simulations} 
 
The above analyzed simulation at $T<1$~K data is fully deterministic. Hence, averaging over repeated realizations is not possible for getting better statistics which would be beneficial for computing, e.g., density distributions.  In order to introduce some degree of	 randomness into the simulations, {we run four equivalent simulations at a temperature of $T=300~K$. {Technically, independence of the four simulations is achieved by (i) using a different realization of the thermally equilibrated substrate, and  (ii) shifting the indenter position by 1 \AA\ in an arbitrary direction.}
Subsequently, the \DtoC{} conversion is applied and the data is averaged in every voxel.} 
{Note that strong thermal vibrations at elevated temperatures might impede the DXA analysis and thus the outcome of the \DtoC{} algorithm. In particular, the DXA analysis becomes unreliable at elevated temperatures close to the melting point. However, it might be possible to partially eliminate the random thermal 
displacements of atoms by performing a time averaging of the atomic positions.}

{\begin{figure*}[htbp]
 	\centerline{\includegraphics{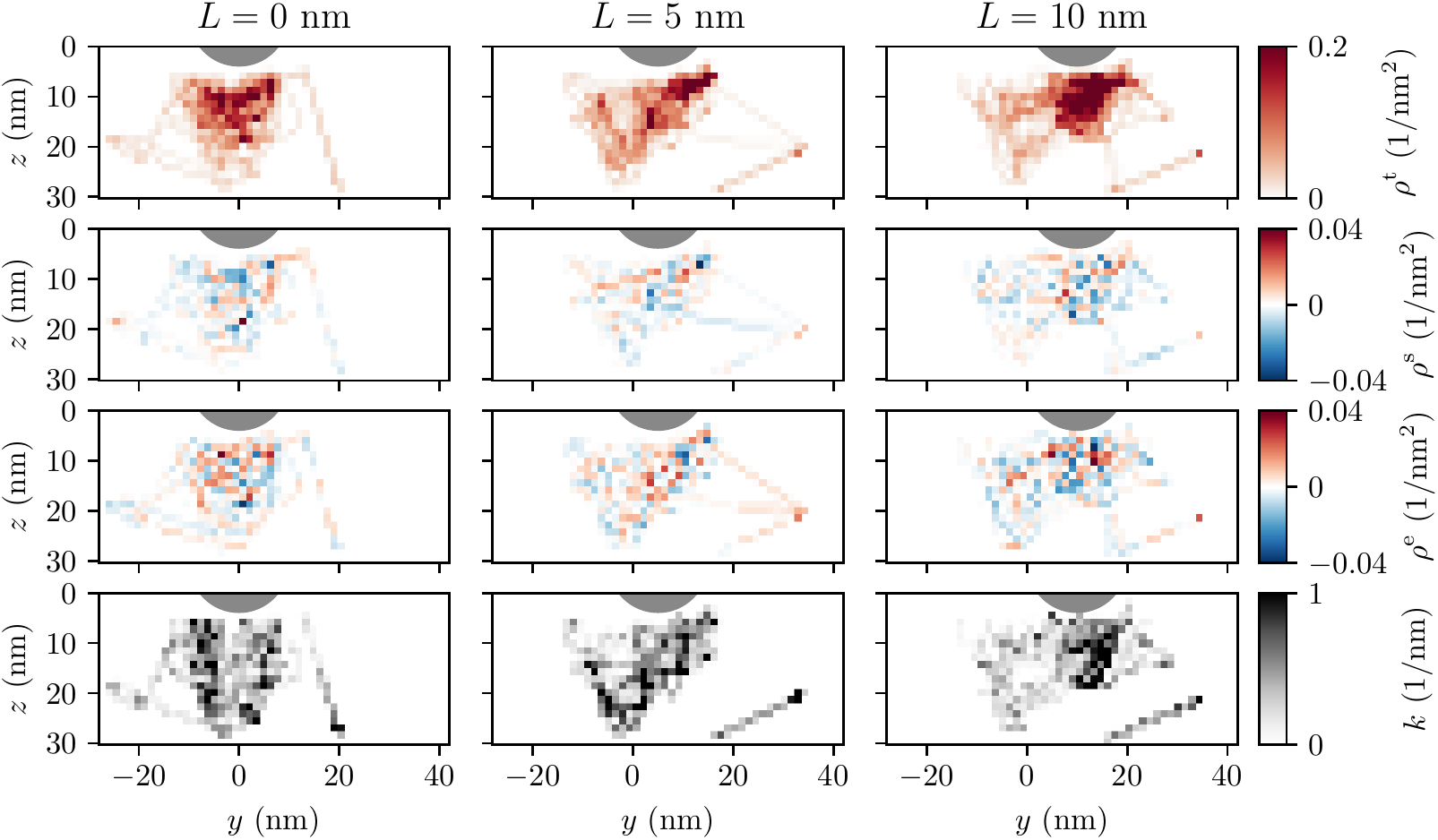}}
 	\caption{{CDD field variables  for scratching the (100) surface. The data from 4 statistically equivalent simulations is averaged, followed by spatial averaging perpendicular to the scratch direction. The position of the indenter is marked in gray.  The edge length of the voxels is $\Delta l\sim 1.5$~nm.}} 
 	\label{f_scratchEvolutionTemp}
\end{figure*}
\begin{figure*}[htbp]
	\centerline{\includegraphics{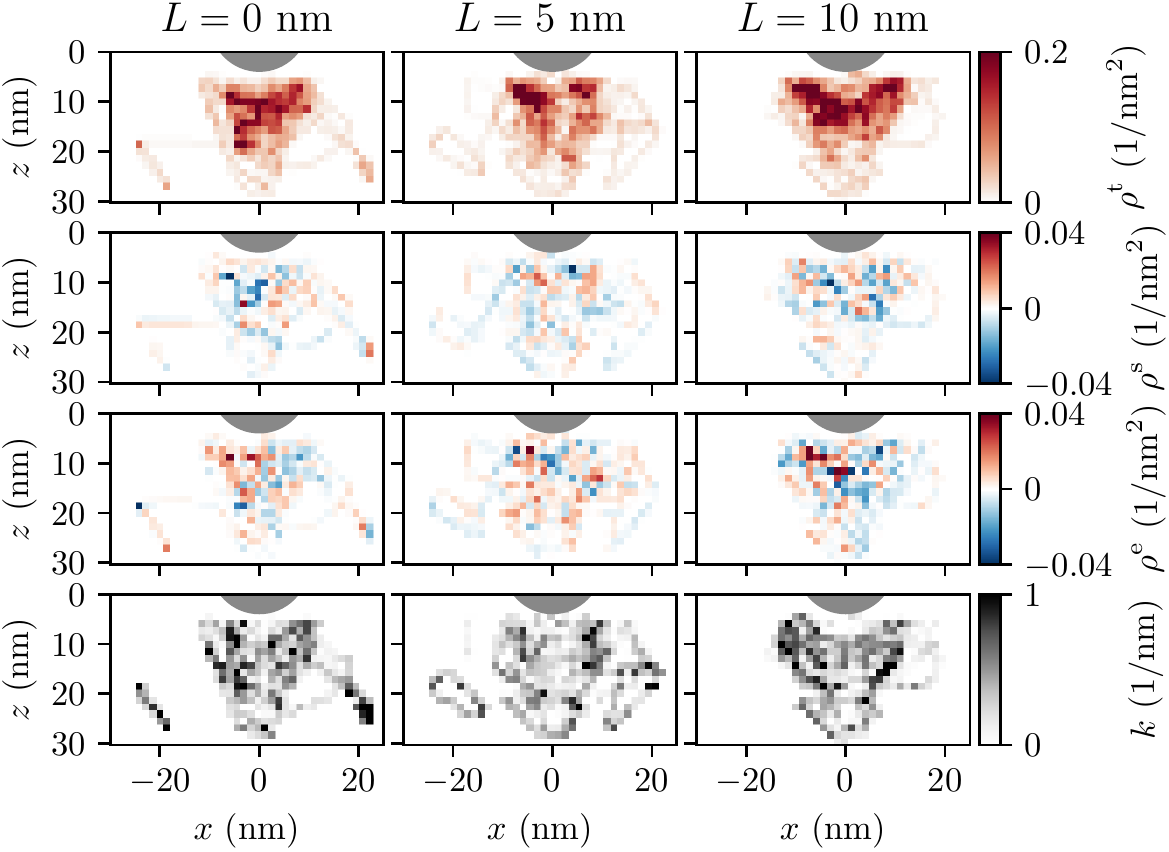}}
	\caption{{CDD field variables for scratching the (100) surface. The data from 4 statistically equivalent simulations is \cmmnt{ensemble} averaged, followed by spatial averaging in scratch direction. The position of the indenter is marked in gray. The edge length of the voxels is $\Delta l\sim 1.5$~nm.}} 
	\label{f_scratchEvolutionTemp2}
\end{figure*}}

Simulations are done for scratching the (100) surface. The \DtoC{} averaging voxels have an edge length of $\approx \SI{3}{nm}$. 
\qfig{f_scratchEvolutionTemp} shows the \cmmnt{ensemble} averaged CDD field variables which are also integrated perpendicular to the scratch direction,
\qfig{f_scratchEvolutionTemp2} shows the same \cmmnt{ensemble} averaged data additionally averaged along the lateral direction. 
The snapshots in \qfig{f_scratchEvolutionTemp2} display a qualitatively similar behavior to the data for $T<1$~K but with less fluctuations due 
to the averaging. The dislocations accumulate at the scratch front and result in regions of high total dislocation density $\rhot$ which, at 
the same time, have a high line curvature $k$. 
The GND density remains small for the temperature considered in this work (approximately 20~\% of $\rhot$). Note that the dislocation density 
for the snapshots averaged along the scratch direction is slightly asymmetric and dislocations mainly pile-up to the left of the indenter. 
Increasing the number of simulations over which averages are computed would remedy this behavior. For the data shown, the asymmetry of total 
density also has an influence on screw and edge GNDs, which are in this region also more pronounced, while the distribution of the curvature 
is almost homogeneous for the snapshots averaged along $y$. 
 

 
\begin{figure}
  \centerline{\includegraphics[width=0.5\textwidth]{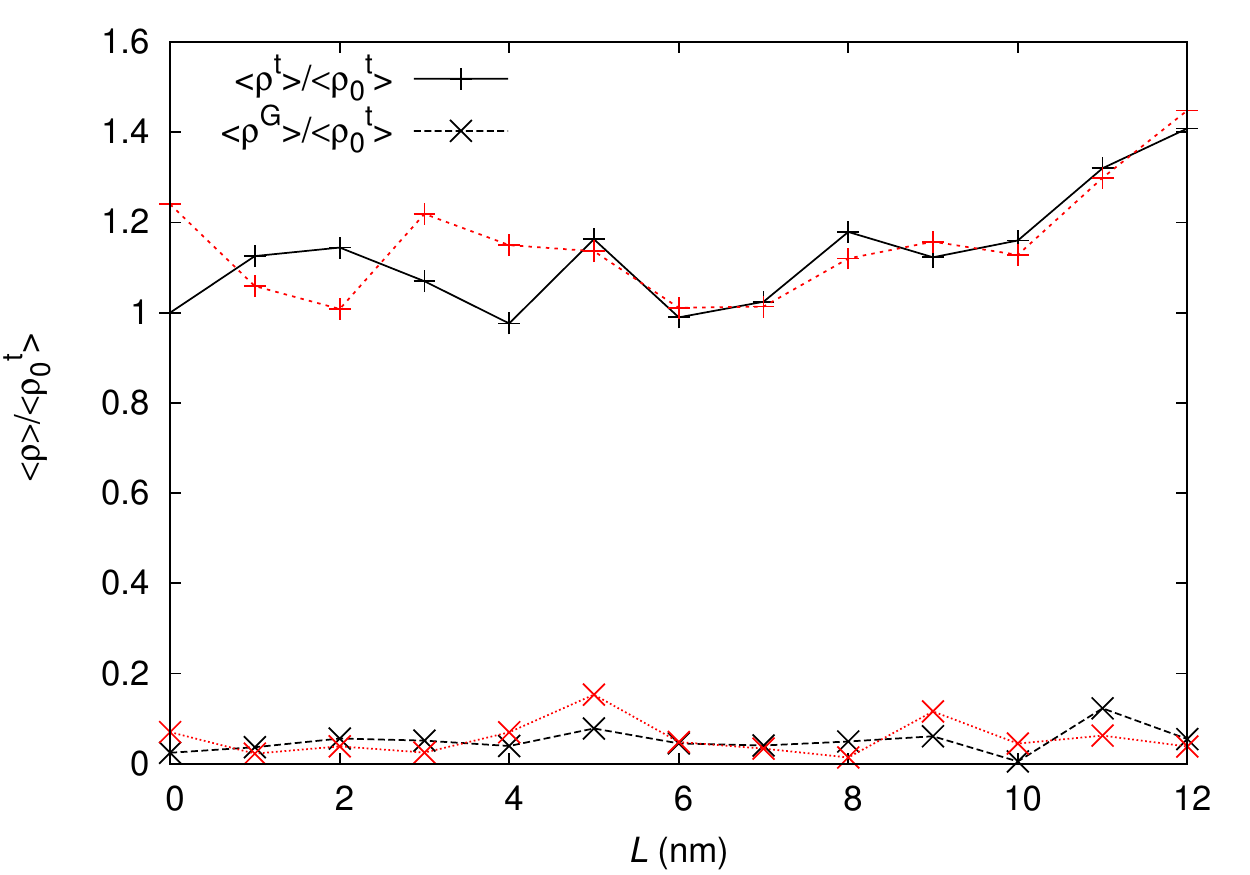}}
 \caption{\label{f_densTemp}
 	{Relative total dislocation density ${\langle \rhot \rangle} / {\langle \rhot_0 \rangle}$ and GND density ${\langle \rhoG \rangle} / {\langle  \rhot_0 \rangle}$, 
  versus scratching length $L$ averaged over 4 different simulations at temperature $T=300~$K where ${\langle \rhot_0 \rangle}$ is the total dislocation density at $L=0$. Scratch of the (100) and (110) surface is marked in black and red, respectively.}} 
\end{figure}

\begin{figure}
  \centerline{\includegraphics[width=0.5\textwidth]{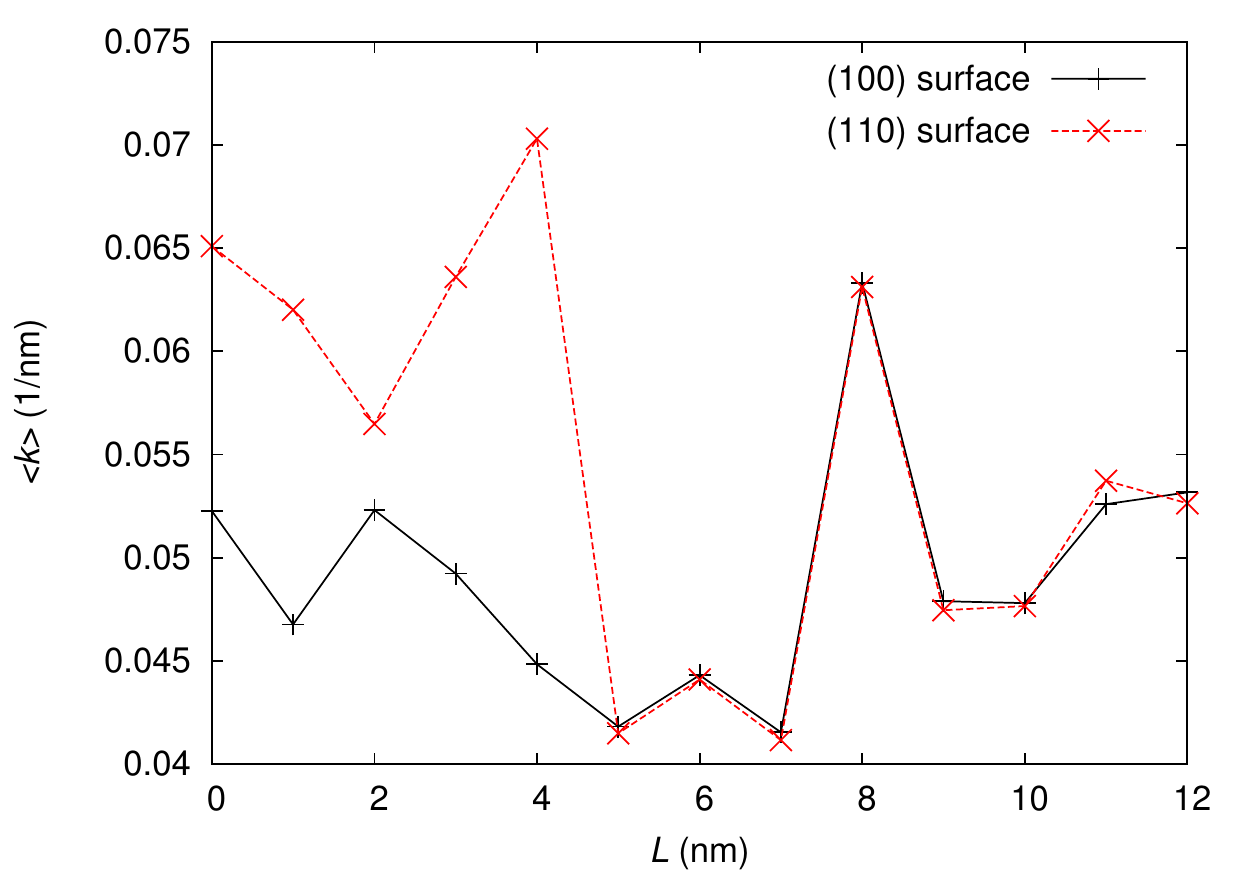}}
 \caption{\label{f_curvTemp}
 	{Curvature $\langle k \rangle$ versus scratching length $L$ averaged over 4 different simulations at temperature $T=300~$K. Scratch of the (100) and (110) surface is marked in black and red, respectively.}} 
\end{figure}

Is the evolution of average density and average curvature found for the single simulation $T<1$~K in \qfig{f_dens} and \qfig{f_curv} typical 
behavior? Comparing those data to the \cmmnt{ensemble} averaged data in \qfig{f_densTemp} and \qfig{f_curvTemp} in fact shows similar behavior in some 
aspects.
The dislocation density $\rhot$ {slightly} increases with scratching length for both orientations while the line length of GNDs 
remains roughly zero, in analogy to the behavior at $T<1$~K. For the \cmmnt{ensemble} averaged simulations, though, the 
difference between the (100) and (110) surfaces seem to become much smaller, suggesting that the large difference for the $T<1$K simulation
might not have a large significance.

%
In \qfig{f_curvTemp} we observe a {rather constant curvature for scratching along (100) and (110). The data strongly fluctuates due to 
the elevated temperature.}

\section{Summary}

In this work we used the recently introduced \emph{discrete-to-continuum (D2C)} method to obtain detailed information about
the microstructure during nanoscratching in bcc iron. 
The data is represented by a number of density and density-like fields containing detailed information about properties of the dislocation microstructure
which cannot directly be captured by atomistic data.
By characterizing the curvature nucleation rate of dislocation loops
versus scratching length we find that for large scratching lengths the nucleation of dislocations is accommodated by annihilation processes leading to an almost constant averaged curvature.
Our data shows a size effect: With increasing scratching length the number of
SSDs increases but the number of GNDs stays constant resulting in reduced von Mises stress. 

\begin{acknowledgments}
N.G. and S.S. gratefully acknowledge financial support from the Deutsche
Forschungsgemeinschaft (DFG) through Research Unit FOR1650 `Dislocation-based
Plasticity' (DFG grant SA 2292/1-2) and I.A.A. and H.M.U. acknowledge support by the Deutsche Forschungsgemeinschaft via the Sonderforschungsbereich 926. Simulations were performed at the High Performance Cluster Elwetritsch (RHRK, TU Kaiserslautern, Germany).
\end{acknowledgments}

\clearpage
\bibliography{scratching}

\newpage \clearpage

\end{document}